\documentclass[12pt]{iopart}

\usepackage{color,colordvi,graphicx,pstcol}

\usepackage{graphicx}
\usepackage{iopams,setstack}
\usepackage[square,comma,numbers,sort&compress]{natbib}
\newcommand{\element}[3]{\langle #1|#2|#3\rangle}

\newcommand{\ket}[1]{|#1\rangle}
\newcommand{\ThreeJ}[6]{\left(\begin{array}{ccc}
#1&#2&#3\\
#4&#5&#6\\
\end{array}\right)}
\newcommand{\SixJ}[6]{\left\{\begin{array}{ccc}
#1&#2&#3\\
#4&#5&#6\\
\end{array}\right\}}
\newcommand{\Hund}{Hund's case (c)}
\newcommand{\sigv}{\hat{\sigma}_{\nu}}
\newcommand{\wfzerog}[9]{\ket{#1#2\frac{#3}{2},\frac{#4}{2};#5#6\frac{#7}{2},-\frac{#8}{2};0_g}#9\ket{#1#2\frac{#3}{2},-\frac{#4}{2};#5#6\frac{#7}{2},\frac{#8}{2};0_g}}
\newcommand{\wfzerogZ}[8]{\ket{#1#2\frac{#3}{2},\frac{#4}{2};#5#6\frac{#7}{2},-\frac{#8}{2};0_g}}
\begin{document}

\input{epsf}
% \include{epsf}
%\draft

\title{Rubidium Rydberg macrodimers}
\author{Nolan Samboy and Robin C\^{o}t\'{e}}
%\author{Robin C\^{o}t\'{e}}

\address{Physics Department, University of Connecticut,
             2152 Hillside Rd., Storrs, CT 06269-3046}
\ead{rcote@phys.uconn.edu}
%\date{\today}

\begin{abstract}

We explore long-range interactions between two 
atoms excited into high principal quantum number $n$ Rydberg states,
and present calculated  potential energy curves
for various symmetries of doubly excited $ns$ and $np$ rubidium
atoms. We show that the potential curves for these symmetries
exhibit deep ($\sim$ GHz) potential wells, 
which can support very extended ($\sim$ $\mu$m) 
bound vibrational states (\textit{macrodimers}). 
We present $n$-scaling relations for both the 
depth $D_e$ of the wells and the equilibrium separations 
$R_e$ of these macrodimers, and explore their response to small electric fields 
and stability with respect to predissociation.
Finally, we present a scheme to form and study
these macrodimers \textit{via} photoassociation, and show how one
can probe the various $\ell$-character of the potential wells.
\end{abstract}

\pacs{32.80.Rm, 03.67.Lx, 32.80.Pj, 34.20.Cf}

%\pacs{33.80.-b, 31.50.-x, 34.20.-b, 34.50.-s
%}
\maketitle

\section{Introduction}
\label{sec:intro}

Rydberg atoms have long been studied because 
of their peculiar properties such as long lifetimes, 
large cross sections, and very large polarizabilities~\cite{Gallagher}.
These exaggerated properties lead to strong interactions between 
the Rydberg atoms, which have been experimentally detected in 
recent years~\cite{Anderson,Mourachko}. Such strong Rydberg-Rydberg 
interactions have fueled a growing interest in the 
field of quantum computing, and over the past decade, 
their application for quantum information processing, such as fast 
quantum gates~\cite{jaksch00,grangier02}, or quantum 
random walks~\cite{cote-qrw} have been proposed.
Also of particular interest is the excitation blockade 
effect~\cite{lukin01}, where one Rydberg atom actually prevents 
the excitation of other nearby atoms in an ultracold 
sample~\cite{tong04,singer04,Liebisch,vogt06,Heidemann08}. 
This phenomenon was recently observed in 
microtraps~\cite{grangier08,saffman09} and a C-NOT gate was
implemented using the behavior~\cite{Saffman10}.

Another active area of research with Rydberg atoms is the predicted existence of 
long-range ``exotic molecules''. In one scenario, one atom remains 
in its ground state, while another atom is excited to a Rydberg state. 
The most famous examples of this type of interaction are the 
\textit{trilobite} and~\textit{butterfly} states, 
so-called because of the resemblence of their respective wave functions 
to these creatures. The theoretical framework for such interactions 
was first proposed in~\cite{trilobites}, but were not observed until 
more recently in~\cite{pfau}. The second type of long-range interaction 
is predicted to occur when both atoms are excited to Rydberg atoms. 
In~\cite{macro-old}, it was first predicted that 
weakly bound \textit{macrodimers} could be formed from the induced 
Van der Waals interactions of two such excited atoms. However, more 
recent work~\cite{Samboy} has shown that larger, more stable
macrodimers can be formed from the strong mixing between $\ell$-characters 
of various Rydberg states. Recent measurements have shown
signatures of such macrodimers in spectra of cesium Rydberg samples \cite{shaffer-NPHYS}.

In this article, we present long-range potential energy curves 
corresponding to the interaction between pairs of rubidium atoms excited to 
$ns$ and $np$ Rydberg states. 
In general, Rydberg-Rydberg interactions will only mix 
states that share the same molecular symmetry~\cite{Marinescu97, Jovica}. 
Thus, only common symmetries between the excited Rydberg molecular state and the state to 
which it is most strongly coupled are relevant.
For rubidium, the doubly excited $ns$ atom pair is most strongly coupled to the 
$np+(n-1)p$ asymptote, while the doubly excited $np$ atom pair is most strongly
coupled to the $ns+(n+1)s$ asymptote. 
Since all $ss'$ states have $m_j=\pm\frac{1}{2}$, the only common symmetries with
any $pp'$ state are $\Omega\equiv |m_{j_1}+m_{j_2}|=0$, 1.
In this manner, we find that the relevant symmetries for the doubly excited $ns$ and 
$np$ asymptotes are $0_g^+$, $0_u^-$ and $1_u$.
We analyze all three cases for both pairs and show that potential wells exist for all of them.
We also describe in detail properties of the bound levels within each 
well.

The paper is arranged as follows: in \S\ref{sec:MolCurves}, we review 
how to build the basis states used to compute the 
potential energy curves at long-range, and describe the existence
of potential wells for certain asymptotes. In \S\ref{sec:EF}, we investigate 
the effects of small external electric fields on the potential curves, 
and in \S\ref{sec:scaling}, we discuss the scaling of the wells with principal 
quantum number $n$. Finally, in \S\ref{sec:Lifetimes}, we calculate
bound levels supported in those wells, and estimate their lifetimes. We also
outline how photoassociation could be used to form and probe macrodimers. This
is followed by concluding remarks in \S\ref{sec:conc}.

%%%%%%%%%%%%%%%%%%%%%%%%%%%%%%%%%%%%%%%%%%%%%%%%%%%%%%%%%%%%%%%%%%%%%%%%%%%%%%%%%%%%%%%%%%%
\section{Molecular Curves}
\label{sec:MolCurves}
\subsection{Basis States}
\label{subs:basis}

In this section, we review the general theory for calculating the 
interaction potential curves. These curves are calculated by 
diagonalizing the interaction Hamiltonian in the Hund's case (c)
basis set, which is appropriate when the spin-orbit coupling 
becomes significant and fine structure cannot be ignored, as is 
the case here.

We first consider two free Rydberg atoms 
in states $|a\rangle\equiv|n,\ell,j,m_j\rangle$ and 
$|a'\rangle\equiv|n',\ell',j',m'_j\rangle$, where
$n$ is the principal quantum number, $\ell$ the orbital angular 
momentum, and $m_j$ is the projection of the total angular 
momentum $\vec{j}=\vec{\ell}+\vec{s}$ onto a quantization 
axis (chosen in the $z$-direction for convenience).
The long-range \Hund\hspace{1pt} basis states are constructed 
as follows:
\begin{equation}
  |a;a';\Omega_{g/u}\rangle \sim |a\rangle_1 |a'\rangle_2 
    - p(-1)^{\ell+\ell'}|a'\rangle_1 |a\rangle_2,
\label{eq:basis}
\end{equation}
where $\Omega = m_j + m'_j$ is the projection of the total angular momentum 
on the molecular axis and is conserved. The quantum number $p$ describes the
symmetry property under inversion and is 1($-$1) for $g(u)$ states.

For $\Omega=0$, we need to additionally account for the reflection 
through a plane containing the internuclear axis.  Such a reflection 
will either leave the wave function unaffected or it will change the 
sign of the wave function. We distinguish between these symmetric
and antisymmetric states under the reflection operator $\sigv$ as follows:
\begin{equation}
\label{eq:refl}
\ket{0^{\pm}_{g/u}}=\frac{1\pm\sigv}{\sqrt{2}} \ket{0_{g/u}}\, ,
\end{equation}
where $\sigv$ behaves according to the following rules~\cite{Bernath,Brown}:
\begin{eqnarray}
\label{eq:refl-rules}
\sigv\ket{\Lambda}&=(-1)^{\Lambda}\ket{-\Lambda}\\
\sigv\ket{S,M_S}&=(-1)^{S-M_S}\ket{S,-M_S}\, .
\end{eqnarray}

References~\cite{Jovica} and~\cite{Jovica-Potentials} give the 
technical details for determining which states comprise the basis 
of the $np+np$ rubidium asymptote. Although the procedure to find 
the basis states for different molecular asymptotes, 
such as $ns+ns$, $np+np$, or $nd+nd$, is the same, 
the states making up these basis sets, in general, will be different. 
We do not review the procedure for building the basis sets
here, but we note that all relevant (\textit{i.e.} strongly coupled) 
molecular asymptotes within the vicinity of the asymptotic 
doubly-excited Rydberg state being considered are included in each respective basis set.
We again note here that
because doubly excited $ns$ ($np$) rubidium atoms are most strongly 
coupled to $pp'$ ($ss'$) states and because only common symmetries of such Rydberg states 
are allowed to mix, the relevant symmetries for the $ns+ns$ and $np+np$ asymptotic Rydberg
states that we consider are $0_g^+$, $0_u^-$ and $1_u$.
As an example, Table~\ref{tab:basis-Rb} lists the basis set for the $0_g^+$ symmetry near the 
Rb $70p+70p$ molecular asymptote.

%%%%%%%%%%%%%%%%%%%%%%%%%%%%%%%%%%
\begin{table}
   \caption{Asymptotic $0_g^+$ molecular states included in the Rb $70p+70p$ basis set, which
            diagonalize the interaction Hamiltonian (see text). The basis states have been 
            symmetrized with respect to the reflection operator~\eref{eq:refl} and
            each $\ket{a_1;a_2;0_g}$ 
            state is defined by equation~\eref{eq:basis}.}
%   \begin{indented} 
    \centering          
   \resizebox{6in}{!}{ 
%   \item[]        
   \begin{tabular}{ll}
   \br
   \vspace{2pt}
   $\frac{1}{\sqrt{2}}\left\{\wfzerog{70}{s}{1}{1}{71}{s}{1}{1}{-}\right\}$ &
   $\frac{1}{\sqrt{2}}\left\{\wfzerog{68}{d}{3}{1}{71}{s}{1}{1}{+}\right\}$\\
   \vspace{3pt}
   $\wfzerogZ{70}{p}{3}{3}{70}{p}{3}{3}$ &
   $\frac{1}{\sqrt{2}}\left\{\wfzerog{68}{d}{5}{1}{71}{s}{1}{1}{-}\right\}$\\
   \vspace{3pt}
   $\wfzerogZ{70}{p}{3}{1}{70}{p}{3}{1}$ &
   $\frac{1}{\sqrt{2}}\left\{\wfzerog{67}{d}{3}{1}{72}{s}{1}{1}{+}\right\}$\\
   \vspace{3pt}   
   $\frac{1}{\sqrt{2}}\left\{\wfzerog{70}{p}{3}{1}{70}{p}{1}{1}{+}\right\}$ &
   $\frac{1}{\sqrt{2}}\left\{\wfzerog{67}{d}{5}{1}{72}{s}{1}{1}{-}\right\}$\\ 
   \vspace{3pt}  
   $\wfzerogZ{70}{p}{1}{1}{70}{p}{1}{1}$ &
   $\frac{1}{\sqrt{2}}\left\{\wfzerog{70}{d}{3}{1}{69}{s}{1}{1}{+}\right\}$\\
   \vspace{3pt}   
   $\frac{1}{\sqrt{2}}\left\{\wfzerog{69}{p}{3}{3}{71}{p}{3}{3}{-}\right\}$ &
   $\frac{1}{\sqrt{2}}\left\{\wfzerog{70}{d}{5}{1}{69}{s}{1}{1}{-}\right\}$\\ 
   \vspace{3pt}  
   $\frac{1}{\sqrt{2}}\left\{\wfzerog{69}{p}{3}{1}{71}{p}{3}{1}{-}\right\}$ &
   $\frac{1}{\sqrt{2}}\left\{\wfzerog{68}{s}{1}{1}{73}{s}{1}{1}{-}\right\}$\\    
   \vspace{3pt}  
   $\frac{1}{\sqrt{2}}\left\{\wfzerog{69}{p}{3}{1}{71}{p}{1}{1}{+}\right\}$ &
   $\frac{1}{\sqrt{2}}\left\{\wfzerog{67}{f}{5}{1}{70}{p}{1}{1}{-}\right\}$\\
   \vspace{3pt}   
   $\frac{1}{\sqrt{2}}\left\{\wfzerog{69}{p}{1}{1}{71}{p}{3}{1}{+}\right\}$ &
   $\frac{1}{\sqrt{2}}\left\{\wfzerog{67}{f}{5}{1}{70}{p}{3}{1}{+}\right\}$\\ 
   \vspace{3pt}  
   $\frac{1}{\sqrt{2}}\left\{\wfzerog{69}{p}{1}{1}{71}{p}{1}{1}{-}\right\}$ &
   $\frac{1}{\sqrt{2}}\left\{\wfzerog{67}{f}{5}{3}{70}{p}{3}{3}{+}\right\}$\\
   \vspace{3pt}   
   $\frac{1}{\sqrt{2}}\left\{\wfzerog{69}{s}{1}{1}{72}{s}{1}{1}{-}\right\}$ &
   $\frac{1}{\sqrt{2}}\left\{\wfzerog{67}{f}{7}{1}{70}{p}{1}{1}{+}\right\}$\\
   \vspace{3pt}   
   $\frac{1}{\sqrt{2}}\left\{\wfzerog{68}{p}{3}{3}{72}{p}{3}{3}{-}\right\}$ &
   $\frac{1}{\sqrt{2}}\left\{\wfzerog{67}{f}{7}{1}{70}{p}{3}{1}{-}\right\}$\\ 
   \vspace{3pt}   
   $\frac{1}{\sqrt{2}}\left\{\wfzerog{68}{p}{3}{1}{72}{p}{3}{1}{-}\right\}$ &
   $\frac{1}{\sqrt{2}}\left\{\wfzerog{67}{f}{7}{3}{70}{p}{3}{3}{-}\right\}$\\       
   \vspace{3pt}   
   $\frac{1}{\sqrt{2}}\left\{\wfzerog{68}{p}{3}{1}{72}{p}{1}{1}{+}\right\}$ &
   $\frac{1}{\sqrt{2}}\left\{\wfzerog{68}{f}{5}{1}{69}{p}{1}{1}{-}\right\}$\\
   \vspace{3pt}   
   $\frac{1}{\sqrt{2}}\left\{\wfzerog{68}{p}{1}{1}{72}{p}{3}{1}{+}\right\}$ &
   $\frac{1}{\sqrt{2}}\left\{\wfzerog{68}{f}{5}{1}{69}{p}{3}{1}{+}\right\}$\\
   \vspace{3pt}   
   $\frac{1}{\sqrt{2}}\left\{\wfzerog{68}{p}{1}{1}{72}{p}{1}{1}{-}\right\}$ &
   $\frac{1}{\sqrt{2}}\left\{\wfzerog{68}{f}{5}{3}{69}{p}{3}{3}{+}\right\}$\\
   \vspace{3pt}
   $\frac{1}{\sqrt{2}}\left\{\wfzerog{69}{d}{3}{1}{70}{s}{1}{1}{+}\right\}$ &
   $\frac{1}{\sqrt{2}}\left\{\wfzerog{68}{f}{7}{1}{69}{p}{1}{1}{+}\right\}$\\
   \vspace{3pt}
   $\frac{1}{\sqrt{2}}\left\{\wfzerog{69}{d}{5}{1}{70}{s}{1}{1}{-}\right\}$ &
   $\frac{1}{\sqrt{2}}\left\{\wfzerog{68}{f}{7}{1}{69}{p}{3}{1}{-}\right\}$\\ 
                                                                             &
   $\frac{1}{\sqrt{2}}\left\{\wfzerog{68}{f}{7}{3}{69}{p}{3}{3}{-}\right\}$\\
   \br
   \end{tabular}
   }
%\end{indented}
\label{tab:basis-Rb}
\end{table}
%%%%%%%%%%%%%%%%%%%%%%%%%%%%%%%%%%%%%%%%%%%%%%%%%%%%%%%%%%%%%%%%%%%%%%%%%%%%%%%%%%%%%%%

\subsection{Long-range Interactions}
\label{subs:longrange}

The interaction matrix we consider consists of both the long-range 
Rydberg-Rydberg interaction and the atomic fine structure. Here, 
``long-range'' refers to the case where no electron exchange takes
place {\it i.e.} the electronic clouds about both nuclei do not overlap. 
This  occurs when the distance $R$ between the two nuclei is 
greater than the LeRoy Radius~\cite{LeRoy}:  
\begin{equation}
R_{LR} = 2\left[\element{n_1\ell_1}{r^2}{n_1\ell_1}^{1/2} 
        + \element{n_2\ell_2}{r^2}{n_2\ell_2}^{1/2}\right]\;.
\label{eq:LeRoy}
\end{equation}
When the distance between the two atoms is larger than 
$R_{LR}$, 
the interaction between them is described by the residual Coulomb potential
between two non-overlapping charge distributions~\cite{Dalgarno}, which can be 
truncated to give only the dipole-dipole ($V_d$) 
and quadrupole-quadrupole ($V_q$) terms. 
For two atoms lying along the $z$-axis, the $V_d$ and $V_q$ terms can be simplified to 
give~\cite{Marinescu95}:
\begin{equation}
V_L(R) = - \frac{(-1)^{\ell}4\pi r_1^Lr_2^L}{\hat{L}R^{\hat{L}}} 
        \sum_m{B_{2\ell}^{L+m}Y_L^m(\hat{r}_1)Y_L^{-m}(\hat{r}_2)}.
\label{eq:RydPot}
\end{equation}
Here, $L=1(2)$ for dipolar (quadrupolar) interactions, 
$B_n^k \equiv \frac{n!}{k!(n-k)!}$ is the binomial coefficient, 
$\vec{r}_i$ is the position of electron $i$ from its center, 
and $\hat{L} \equiv 2L+1$.

Because the molecular basis states are linear combinations of the 
atomic states determined through symmetry considerations, 
each matrix element will actually be a sum of multiple interactions, 
\textit{i.e.} 
\begin{eqnarray}
\element{a;a';\Omega_{g/u}}{V_L}{b;b';\Omega_{g/u}}& = & \element{a;a'}{V_L}{b;b'}
          -p_a(-1)^{\ell_a+\ell'_a}\element{a';a}{V_L}{b;b'} \nonumber \\ &  &
          -p_b(-1)^{\ell_b+\ell'_b}\element{a;a'}{V_L}{b';b} \nonumber \\ &  &
          +p_a p_b (-1)^{\ell_a+\ell'_a+\ell_b+\ell'_b}\element{a';a}{V_L}{b';b} \;,
\label{eq:VintExp}
\end{eqnarray}
where $|a;a'\rangle \equiv |a\rangle_1|a'\rangle_2$ and so on.
An analytical expression for the long-range interactions 
is obtained using angular momentum algebra in terms of $3$-$j$ and $6$-$j$ symbols:
\begin{eqnarray}
\label{eq:MtxEl}
&\element{1;2}{V_L(R)}{3;4}=(-1)^{L-1-\Omega+j_{\rm tot}}\sqrt{\hat{\ell_1}\hat{\ell_2}\hat{\ell_3}\hat{\ell_4}\hat{j_1}\hat{j_2}\hat{j_3}\hat{j_4}} \, \, \, \, \frac{\mathcal{R}_{13}^L\mathcal{R}_{24}^L}{R^{2L+1}}\nonumber\\
&\times\ThreeJ{\ell_1}{L}{\ell_3}{0}{0}{0}\ThreeJ{\ell_2}{L}{\ell_4}{0}{0}{0}\SixJ{j_1}{L}{j_3}{\ell_3}{\frac{1}{2}}{\ell_1}\SixJ{j_2}{L}{j_4}{\ell_4}{\frac{1}{2}}{\ell_2}\\
&\times\sum_{m=-L}^{L}B_{2L}^{L+m}\ThreeJ{j_1}{L}{j_3}{-m_{j_1}}{m}{m_{j_3}}\ThreeJ{j_2}{L}{j_4}{-\Omega+m_{j_1}}{-m}{\Omega-m_{j_3}}\nonumber \, ,
\end{eqnarray}
where $j_{\rm tot} \equiv \sum_{i=1}^4{j_i}$, $\hat{\ell_i}=2\ell_i+1,\hat{j_i}=2j_i+1$, 
and $\mathcal{R}_{ij}^L=\element{i}{r^L}{j}$ is the radial matrix element.  
For $\ket{1;2}=\ket{3;4}$, the matrix element is given by 
\eref{eq:MtxEl} plus the sum of the two atomic asymptotic Rydberg energy values.  
That is:
\begin{equation}
\element{1;2}{V_L(R)}{1;2} = \mbox{\eref{eq:MtxEl}} + E_1+E_2\,,
\label{eq:diag}
\end{equation}
with $E_i$ given by 
%$-\displaystyle\frac{R_yc}{(n_i-\delta_\ell)^2}$, 
%where $R_y$ is the Rydberg constant, $c$ is the speed of light,
$-\,\displaystyle\frac{1}{2(n_i-\delta_{\ell})^2}\,\, ,$
where $n_i$ is the principal quantum number and $\delta_\ell$ is the quantum defect 
(values given in~\cite{Li} and~\cite{nfqd}). Since $\Delta \ell=0$ dipole transitions 
are forbidden, only the $L=2$ term of equation~\eref{eq:MtxEl} will contribute
in~\eref{eq:diag}.
%%%%%%%%%%%%%%%%%%%%%%%%%%%%%%
\begin{figure}[t]
	\centering
	 \includegraphics[width=6in]{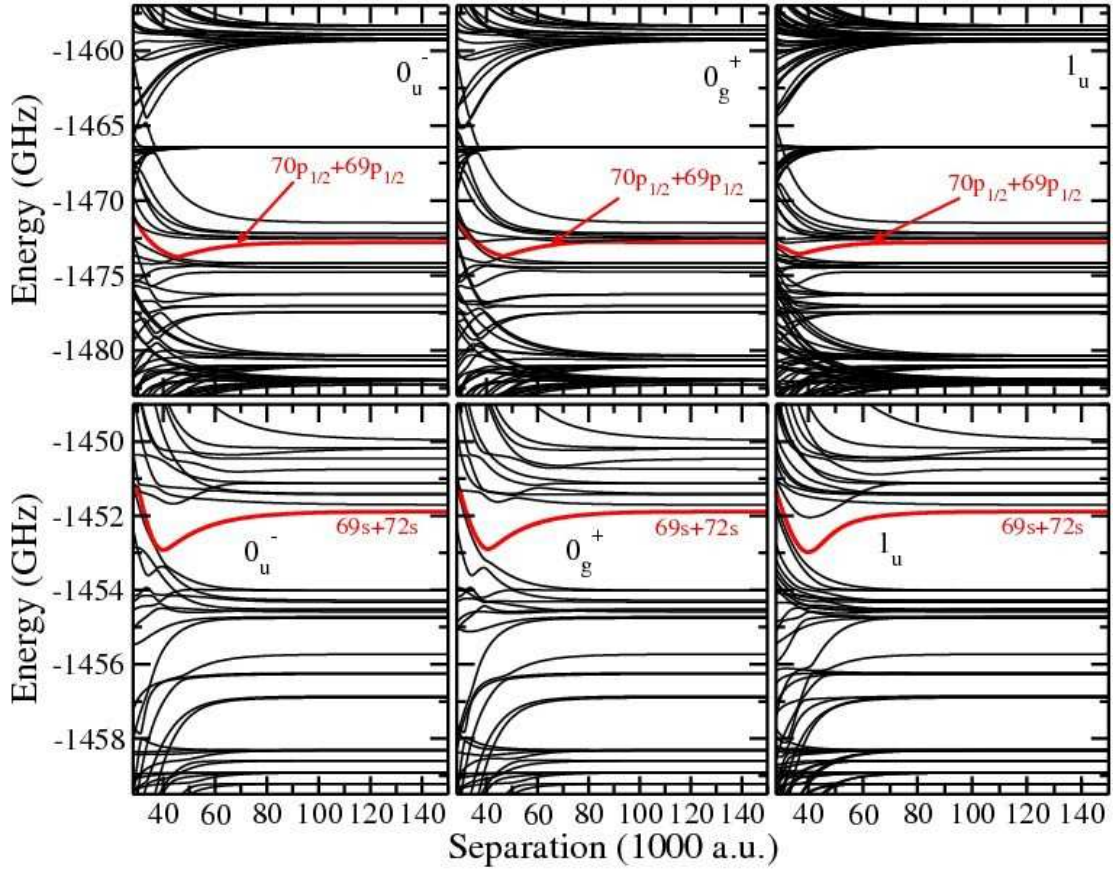}
	\caption{(Color online) Long-range interaction curves for 
                  the $0_u^-$ (left), $0_g^+$ (middle),
                   and $1_u$ (right) symmetries of doubly-excited
	          $ns$ (top row), and $np$ (bottom row) Rb Rydberg atoms near $n=70$.
	          We highlight the potential wells and label their corresponding 
	          asymptotic state. 
	          Note: All energies are measured from the Rb ionization threshold.
	          }		
	\label{fig:70spdALL}
\end{figure}
%%%%%%%%%%%%%%%%%%%%%%%%%%%%%

Figure~\ref{fig:70spdALL} shows the results of diagonalization for the
$0_g^+$, $0_u^-$ and $1_u$ symmetries of the $70s+70s$, and $70p+70p$ 
asymptotes with no background electric field. 
In all of these plots, the energies are measured from the ionization threshold of rubidium.
We see that all three symmetries feature large potential wells and for 
the remainder of this paper, we focus our attention on formation properties 
of bound states within these wells and analyze the stability of these macrodimers.
We note here that we also explored the potential curves near the $nd+nd$ asymptotes,
but no wells were found; hence we do not display those curves here.

%%%%%%%%%%%%%%%%%%%%%%%%%%%%%%%%%%%%%%%%%%%%%%%%%%%%%%%%%
\section{Electric Field Dependence}
\label{sec:EF}

Production and/or detection of macrodimers will rely 
on external electric fields. In addition, since experiments cannot 
completely shield the atoms from undesired stray fields, it is important 
to study their effect on our calculated curves.

Strictly speaking, applying an external electric field $\vec{F}$  
breaks the $D_{\infty h}$ symmetry of homonuclear dimers, 
and consequently, the basis states defined by \eref{eq:basis} would no 
longer be valid. In principle, one then needs to diagonalize the 
interaction matrix in a basis set containing every possible Stark 
state, as was done in~\cite{shaffer}. 
However, since the effects of such an electric field should be adiabatic, 
we assume that the $D_{\infty h}$ symmetry is still approximately valid 
for small electric fields.

\begin{figure}
	\centering
		\includegraphics[width=5.5in]{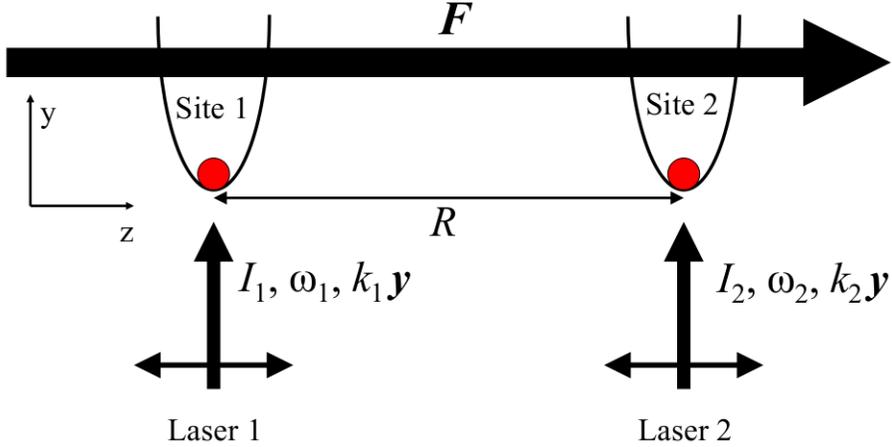}
	\caption{(Color online) Two Rydberg atoms confined within harmonic 
	         traps provided by optical lattice sites.
	         The electric field $\mathbf{F}$ is directed
	         along the $z$-axis, coinciding with the molecular axis of the dimer. 
	         We also indicate in this cartoon that the two excitation lasers 
	         (see section~\ref{subs:PA}) 
	         propagate in the $y$-direction, but the polarization directions of 
	         both lasers are along the $z$-axis.}
	\label{fig:EFscheme}
\end{figure}

%Due to the large size of Rydberg atoms {\bf \color{red} and their large 
%polarizability, they are highly sensitive to electric fields}, we can
%use a semi-classical model for our discussion, where the energy of the electric field is
We consider the effects of such an electric field as a perturbation to the 
original Hamiltonian. 
In general, an applied electric field will define a quantization axis; the molecular axes
of our macrodimers will then be at some random angle to this quantization axis.
In that case, one needs to transform the molecular-fixed frame back 
into the laboratory-fixed frame \cite{krems}.
To simplify our calculations, we assume that the two Rydberg atoms are first 
confined in an optical lattice, such that the quantization axes of the macrodimer 
and the electric field coincide (see figure~\ref{fig:EFscheme}).
Such one-dimensional optical lattices have already been used to experimentally
excite Rydberg atoms from small Bose-Einstein condensates located at individual
sites \cite{arimondo}. We envision a similar one-dimensional optical lattice with
the distance between adjacent (or subsequent) sites adjusted to coincide with 
$R_e$, but containing a single atom per site. The lattice could be switched off during the
Rydberg excitation to allow a cleaner signal.
%%%%%%%%%%%%%%%%%%%%%%%%%%%%%%%%%%%%%%%%%%%%%%%%%%%%
%%%%%%%%%%%%%%%%%%%%%%%%%%%%%%%%%%%%%%%%%%%%%%%%%%%%

\begin{figure}[t]
	\centering \includegraphics[height=3in]{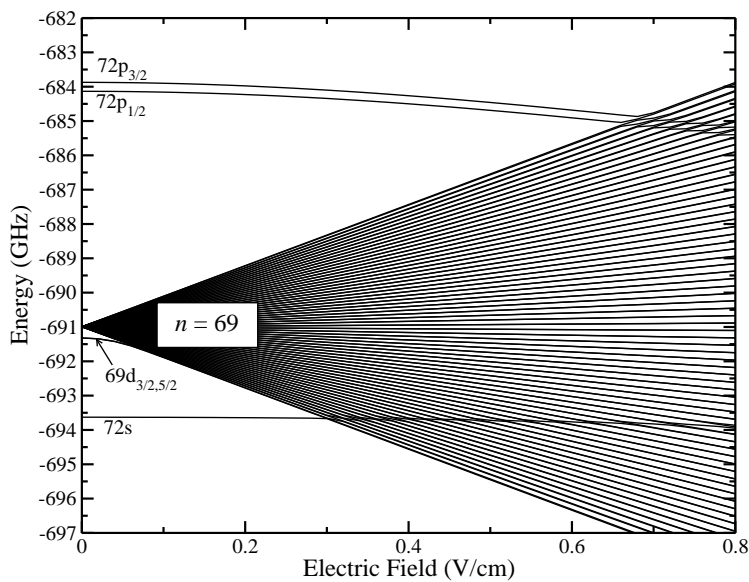}
	\caption{Atomic Rydberg energies vs. the 
                electric field for Rb with $|m_{\ell}|=0$ near $n=69$: 
                the curves labeled by $n=69$ include states with $\ell\geq 3$.
	         }
	\label{fig:Stark}
\end{figure}

For an electric field directed along the $z$-axis, 
the perturbation Hamiltonian for a single atom is given %{in atomic units 
by $Fr\cos{\theta}$, where $F$ is the magnitude of the electric field, 
$r$ is the distance of the valence electron from its nucleus, and 
$\theta$ is the angle between $F$ and $r$.
We express the new eigenstates (Stark or dressed states) of the 
perturbed Hamiltonian as a series expansion using the unperturbed
asymptotic eigenstates, \textit{i.e.}
\begin{equation}
\label{eq:stark}
\ket{ \tilde{a}}_k = \sum_i  b_{k,i} (F) \ket{a_i}_k\,\,,
\end{equation}
where $\ket{a_i}_k \equiv \ket{n_i, \ell_i , j_i, m_j^{(i)} }_k$ 
are the unperturbed (undressed) atomic states of atom $k$, 
and $b_{k,i}(F)$ are field-dependent eigenvectors, resulting from 
diagonalization~\cite{Zimmerman}.
Here, the index $i$ stands for $n_i$, $\ell_i$, and $j_i$. 
In Fig.~\ref{fig:Stark}, we show the Stark map for $|m_{\ell}|=0$ 
near $n=69$. Although the limits of this summation are technically 
$n_i \rightarrow n_{max}$ (where $n_{max}$ is the highest $n$ 
value in the basis) and $\ell_i \rightarrow (n_i-1)$, 
we restrict the summation to $(n-2)<n_i<(n+2)$ and $\ell_i \leq 3$; 
the $b_i(F)$ coefficients are insignificant ($\geq$ 2 orders of 
magnitude smaller) for states lying outside these bounds.
Using the dressed states~\eref{eq:stark} 
in~\eref{eq:basis}, we define the dressed molecular states 
as:
\begin{equation}
   \ket{\tilde{a}}_1\ket{\tilde{a}}_2 = \sum_{ij} b_{1,i}(F)b_{2,j}(F) \ket{a_i}_1\ket{a_j}_2 \,\, .
\label{eq:dressed}
\end{equation}
We then use this basis to redefine the properly symmetrized
dressed molecular basis given in Table~\ref{tab:basis-Rb} and to diagonalize the 
Rydberg-Rydberg interaction matrix.
 
\begin{figure}[h]
	\centering
		\includegraphics[height=3in]{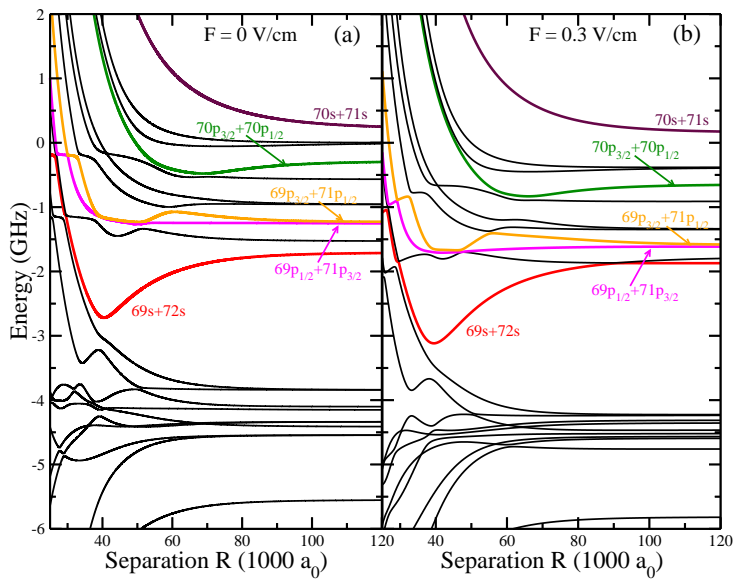}
	\caption{(Color online) $0_g^+$ molecular curves for Rb $70p+70p$: (a) $F=0$, and
	(b) $0.3$ V/cm. The states involved in the well correlated to $69s+72s$
        are identified (see section~\ref{sec:Lifetimes}). 
	The zero-energy is set at the $70p_{3/2}+70p_{3/2}$ asymptote in both plots.}
	\label{fig:EFboth}
\end{figure}

In Fig.~\ref{fig:EFboth}, we illustrate the effect of $\vec{F}$ on the curves
near $70p+70p$ of the $0_g^+$ symmetry, in a side-by-side comparison of 
the curves for $F=0$ and 0.3 V/cm. The atomic Stark states were computed using the method
described in \cite{Zimmerman,macro-old} and the interaction curves for the
``pseudo-symmetry" were obtained by diagonalization of the Rydberg-Rydberg 
interaction matrix in the dressed molecular basis set~\eref{eq:dressed}.
We find some minor differences: the Stark effect is most notable in the
shifting of the potential curves, especially the asymptotic energies. However, 
the relative shape of the curves are only slighty changed and most importantly, 
the large potential well is robust against small electric fields.

%%%%%%%%%%%%%%%%%%%%%%%%%%%%%%%%%%%%%%%%%%%%%
%%%%%%%%%%%%%%%%%%%%%%%%%%%%%%%%%%%%%%%%%%%%%%%%

%%%%%%%%%%%%%%%%%%%%%%%%%%%%%%%%%%%%%%%%%%%%%%%%%%%%%%%%%%%%%%%%%%%%%%%%%%%%%%%%%%%%%%%%%%%%%

\section{Scaling}
\label{sec:scaling}

\begin{figure}[h]
	\centering
		\includegraphics[height=4in]{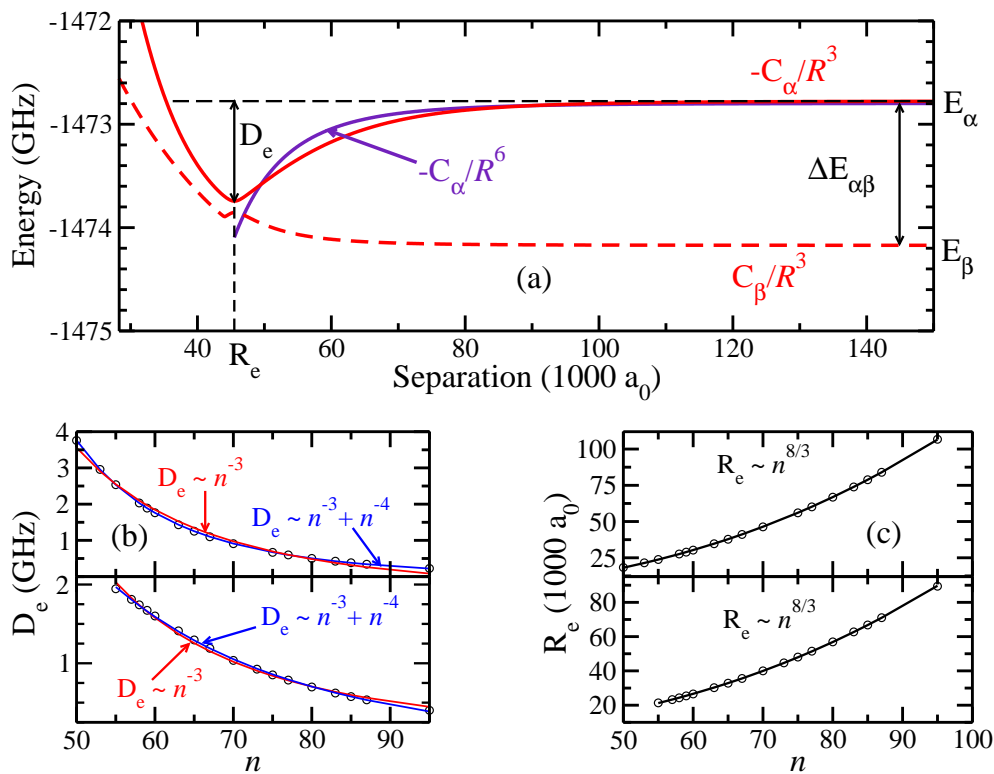}
	\caption{(Color online) (a) Isolated avoided crossing in the $0_u^-$ 
                 symmetry curves of doubly excited $ns$ atoms, which results 
                 in the potential well correlated to the 
                 $69p_{1/2}+70p_{1/2}$ asymptote. We assume the 
                 interactions at the crossing are mostly dipolar in nature (see text) 
                 and label the well depth $D_e$ and the 
	               equilibrium separation $R_e$ at the avoided crossing. We also
	               demonstrate that at long range the curves behave as 
	               $\sim 1/R^6$, but in the $R$-range of the avoided
	               crossing, they behave as $\sim 1/R^3$.
	               (b) Scaling relations for the well 
                 depth $D_e$ vs $n$ for the $0_u^-$ symmetry of $ns+ns$ (top) 
                 and $np+np$ (bottom). (c) Scaling relations
	               for the equilibrium separation $R_e$ vs $n$ for the same 
	               curves as in (b).
	         }
	\label{fig:scaling}
\end{figure}

We focus our attention on the wells correlated to the $(n-1)s+(n+2)s$ asymptote 
near $np_{3/2}+np_{3/2}$, and those correlated to $(n-1)p_{1/2}+n p_{1/2}$ 
near $ns+ns$, for $F=0$. We calculated the curves for a large range of $n$, and
found that the wells follow simple $n$-scaling behaviors (see Fig.~\ref{fig:scaling}).
As will be discussed in \S\ref{sec:Lifetimes}, the wells are 
produced by the mixing of several states with different $\ell$-character. 
The exact mixing takes place mainly {\it via} dipole and quadrupole 
interactions, and occurs between nearby $n$-states: hence the wells 
depend on the combination of multipole interaction between states, and 
the proximity (in energy) of those states.

To derive a simple $n$-scaling behavior for both the depth $D_e$ and 
equilibrium separation $R_e$, we assume that the dipole-dipole coupling 
is the dominant interaction between states, and that the wells are 
formed as a result of an avoided crossing between two potential curves
(see Fig.~\ref{fig:scaling}(a)). As mentioned above, the real situation 
is much more complex, but these assumptions allow for a simple treatment.

%%%%%%%%%%%%%%%%%%%%%%%%%%%%%%%%%%%%%%
%%%%%%%%%%%%%%%%%%%%%%%%%%%%%%%%%%%%%

The energy difference $\Delta E_{\alpha\beta} = E_\alpha - E_\beta$ 
is defined by the difference between the asymptotes of the two crossing states,
$\alpha$ and $\beta$.
Here, $\alpha = n_{\alpha_1}s + n_{\alpha_2}s$, with energy
$2E_\alpha = -(\nu_{\alpha_1}^{-2} + \nu_{\alpha_2}^{-2})$ and 
$\beta = n_{\beta_1}p + n_{\beta_2}p$, with energy 
$2E_\beta = -(\nu_{\beta_1}^{-2} + \nu_{\beta_2}^{-2})$. In both energy terms,
$\nu_{\gamma_i} \equiv n_{\gamma_i} - \delta_{\gamma_i}$ includes
the principal quantum number $n$ and quantum defect $\delta$ of
the appropriate atomic state $\gamma_i$ of atom $i$. Assuming that
the relevant atomic states in a given asymptote are separated by 
$\Delta n$ of the order unity, we can expand the energies as
\begin{equation}
   2E_\alpha = -2 \frac{\Delta n_\alpha}{\nu_{\alpha_1}^3} \makebox[.5in]{ and }
   2E_\beta = -2 \frac{\Delta n_\beta}{\nu_{\beta_1}^3} \; ,
\end{equation}
so that
\begin{eqnarray}
   \Delta E_{\alpha\beta}& \simeq &-\left[ 
      \frac{\Delta n_\alpha}{\nu_{\alpha_1}^3}
    + \frac{\Delta n_\beta}{\nu_{\beta_1}^3} \right] 
      \simeq  - \frac{\Delta n_\alpha}{\nu_{\alpha_1}^3} \left[ 
    1 - \frac{\Delta n_\beta}{\Delta n_\alpha} 
    \left( 1 - 3\frac{\Delta \nu}{\nu_{\alpha_1}} \right) \right] \;, \nonumber \\
    & \equiv & -A \nu_{\alpha_1}^{-3} - B \nu_{\alpha_1}^{-4} \; ,
\end{eqnarray}
where we assume $\nu_{\beta_1} \simeq \nu_{\alpha_1} + \Delta \nu$ 
with $\Delta \nu$ of order one, and $A = (\Delta n_\alpha - 
\Delta n_\beta)$ and $B= 3\Delta \nu \Delta n_\beta$. 
In the cases leading to our wells, $\Delta n_\alpha \simeq \Delta n_\beta$
so that the leading dependence is $\Delta E_{\alpha\beta}\propto n^{-4}$.

>From the sketch depicted in Fig.~\ref{fig:scaling}(a), assuming leading
dipole-dipole interactions, the equilibrium separation $R_e$ occurs at the ``intersection'' of
two attractive and repulsive curves separated by 
$\Delta E_{\alpha \beta}$, {\it i.e.}, 
$\Delta E_{\alpha \beta} - C_{\alpha}/R_e^3 \simeq + C_{\beta}/R_e^3$, which
lead to $R_e \simeq [(C_{\beta}+C_{\alpha}]/\Delta E_{\alpha \beta}]^{1/3}$.
Our assumption that the two crossing curves behave as $\sim 1/R^3$ is valid in the region
of the intersection; at larger values of $R$, however ($R \gtrsim$ 80~000 $a_0$), the curves
behave more like $\sim 1/R^6$ (see figure~\ref{fig:scaling}).
>From the scaling $C_{\beta}+C_{\alpha} \propto n^4$ and 
$\Delta E_{\alpha \beta} \propto n^{-4}$, we obtain $R_e \propto n^{8/3}$. 
As for the dissociation energy $D_e$, it is simply given by
$\Delta E_{\alpha \beta} - C_{\alpha}/R_e^3 \simeq D_e$, which scales as 
$\Delta E_{\alpha \beta}$, {\it i.e.} $D_e \propto n^{-4}$. 
Fig.~\ref{fig:scaling}(b) shows a plot of $D_e$ vs. $n$ for the $0_u^-$ 
symmetry of the $ns+ns$ and $np+np$ asymptotes,  indicating that $D_e$ 
indeed scales more like $\sim n^{-3} + n^{-4}$ (blue curve) than purely 
$n^{-3}$ (red curve). For the same wells, Fig.~\ref{fig:scaling}(c) shows 
that $R_e$ follows the predicted $n^{8/3}$ scaling.
In the interest of space, we do not show plots for 
all of the symmetries highlighted in Fig.~\ref{fig:70spdALL}, 
but we note that we find the same approximate scaling for all other
symmetries.

Although the analytical derivations above give good agreements 
with numerically determined values of $D_e$ and $R_e$, the slight 
discrepencies, especially with the $np+np$ plots, 
reflect the more complex nature of the interactions. For example, 
quadrupole coupling is present in our calculations
(although its effect is generally small). We also point out 
that in the three $np+np$ cases, the formation of each well 
is not clearly given by an avoided crossing of two curves, but 
rather by several interacting curves (see next section). 
Nonetheless, the good agreement depicted in Fig.~\ref{fig:scaling}
indicate that these more complicated interactions act only as 
small corrections.

%%%%%%%%%%%%%%%%%%%%%%%%%%%%%%%%%%%%%%%%%%%%%%%%%%%%%%%%%%%%%%%%%%%%%%%%%%%%

\section{Forming Macrodimers}
\label{sec:Lifetimes}

\subsection{Energy Levels and Lifetimes}
\label{subs:Levels}

The wells identified in Fig.~\ref{fig:70spdALL} support many bound levels.
We list the lowest levels for each well in Table~\ref{tab:Lifetimes}, 
together with the corresponding classical inner and outer turning points.
For those wells around the $n\sim 70$ asymptotes, the deepest energy levels 
are separated by about 1$-$2 MHz, corresponding to oscillation periods of a 
few $\mu$s, rapid enough to allow for several oscillations during the lifetime 
of the Rydberg atoms (roughly a few hundred $\mu$s for $n=70$). These energy
splittings also allow for detection through spectroscopic means. As illustrated
by the values of the turning points in Table~\ref{tab:Lifetimes}, the bound 
levels are very extended, leading to {\it macrodimers} of a few $\mu$m in size.

%%%%%%%%%%%%%%%%%%%%%%%%
\begin{table}
   \caption{Energies of the six deepest bound levels (measured from the
            bottom of the well) and corresponding
            classical turning points for the $0_g^+$, $0_u^-$ and 
            $1_u$ symmetries near doubly excited $ns$ and $np$ Rb Rydberg 
            atoms near $n=70$.  
            }    
%   \centering
   \begin{indented}
   \lineup
   \item[]                
   \begin{tabular}{@{}llllll}
   \br
    Asymptote& Symmetry& $v$ & Energy (MHz) & $R_1$ (a.u.) & $R_2$ (a.u.) \\
   \br
    $ns+ns$ & $0_g^+$ & 0 & \01.035 & 46,137 & 46,538 \\
            &         & 1 & \03.121 & 45,985 & 46,688 \\
            &         & 2 & \05.353 & 45,870 & 46,800 \\
            &         & 3 & \07.477 & 45,780 & 46,888 \\
            &         & 4 & \09.567 & 45,702 & 46,963 \\
            &         & 5 &  11.645 & 45,630 & 47,032 \\
    \mr
    $ns+ns$ & $0_u^-$ & 0 & \01.034 & 45,072 & 45,488 \\
            &         & 1 & \03.161 & 44,913 & 45,650 \\ 
            &         & 2 & \05.262 & 44,803 & 45,761 \\      
            &         & 3 & \07.331 & 44,716 & 45,854 \\
            &         & 4 & \09.368 & 44,639 & 45,934 \\
            &         & 5 &  11.392 & 44,569 & 46,008 \\
    \mr
    $ns+ns$ & $1_u$   & 0 & 0.917 & 36,181 & 36,600 \\
            &         & 1 & 2.699 & 36,038 & 36,755\\
            &         & 2 & 4.467 & 35,938 & 36,868\\
            &         & 3 & 6.240 & 35,859 & 36,963\\
            &         & 4 & 7.987 & 35,789 & 37,046\\
            &         & 5 & 9.747 & 35,730 & 37,121\\       
    \br
    $np+np$ & $0_g^+$ & 0 & 0.831 & 40,228 & 40,679  \\
            &         & 1 & 2.499 & 40,068 & 40,849  \\
            &         & 2 & 4.166 & 39,959 & 40,970  \\
            &         & 3 & 5.824 & 39,870 & 41,068  \\
            &         & 4 & 7.477 & 39,795 & 41,154  \\
            &         & 5 & 9.125 & 39,728 & 41,233  \\
    \mr
    $np+np$ & $0_u^-$ & 0 & 0.800 & 39,753 & 40,212\\
            &         & 1 & 2.414 & 39,590 & 40,381\\
            &         & 2 & 4.023 & 39,479 & 40,509\\
            &         & 3 & 5.631 & 38,390 & 40,610\\
            &         & 4 & 7.234 & 39,312 & 40,699\\
            &         & 5 & 8.832 & 39,244 & 40,778\\
    \mr
    $np+np$ & $1_u$   & 0 & 0.708 & 36,907 & 37,361 \\
            &         & 1 & 2.212 & 36,752 & 37,535\\
            &         & 2 & 3.721 & 36,632 & 37,635\\
            &         & 3 & 5.219 & 36,545 & 37,780\\
            &         & 4 & 6.734 & 36,467 & 37,870\\
            &         & 5 & 8.238 & 36,399 & 37,954\\        
   \br
   \end{tabular}
   \end{indented}
\label{tab:Lifetimes}
\end{table}
%%%%%%%%%%%%%%%%%%%%%%%%%%%%%%%%%%%%%%%%

As described in \S\ref{sec:MolCurves}, the molecular curves 
are a direct result of the $\ell$-mixing occuring between the 
electronic basis states~\eref{eq:basis}: each molecular electronic state 
$|\chi_{\lambda}(R)\rangle$ (corresponding to the potential 
curve $U_\lambda (R)$) is expanded onto the electronic 
basis states, the amount of mixing varying with $R$:
\begin{equation}
   |\chi_\lambda (R) \rangle =\sum_j c_j^{(\lambda)}(R) |j\rangle\, ,
   \label{eq:electronic-state}
\end{equation}
where $c_j^{(\lambda)}(R)$ are the eigenvectors after diagonalization 
for each separation $R$, and $|j\rangle$ the electronic basis states~\eref{eq:basis}.

\begin{figure}[h!]
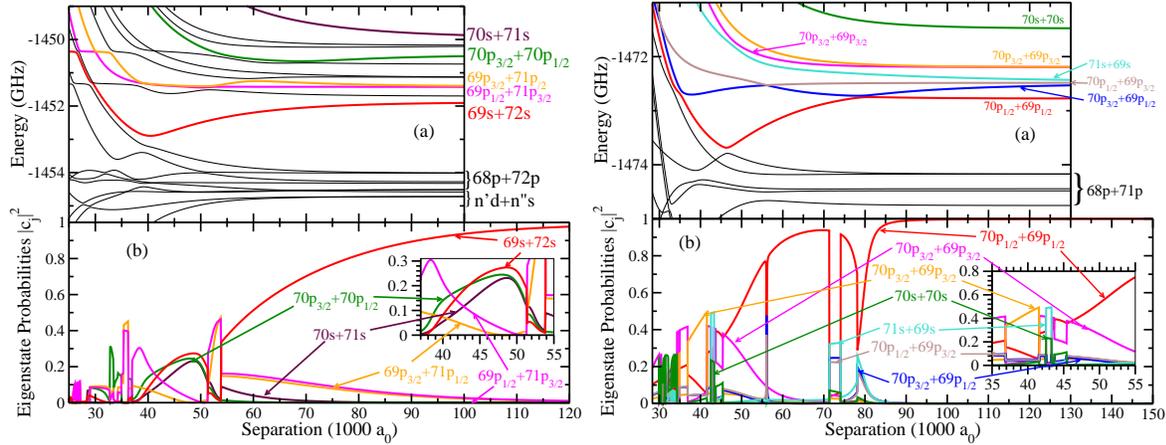

	\centering			
	\includegraphics[width=3.0in]{Curves_eigvex.eps} 
        \includegraphics[width=3.0in]{70sEigvex_2.eps}
		\caption{(Color online)
		         (a) $0_g^+$ symmetry curves of 
                          the rubidium $70p+70p$ molecular asymptote
		          %zeroed at the $70p_{3/2}+70p_{3/2}$ asymptote 
                          (left panel), and the $70s+70s$ molecular asymptote; 
		         % zeroed at the $70s+70s$ asymptote. 
		          both plots are zeroed at the ionization level of rubidium.
		          For both panels, 
		          we highlight the molecular curves corresponding 
                          to the five electronic states contributing the 
                          most to the formation of the well (see text).
		          (b) Composition of the 69$s+$72$s$ well (left panel)
                          and of the $69p_{1/2}+70p_{1/2}$ well (right panel):
                          probabilities $|c_j(R)|^2$ of the electronic states 
                          that contribute the most to the formation
		          of the well vs. the nuclear distance $R$. 
                          Inset: zoom of the inner region.
		          }
	\label{fig:eigvexPLUS}
\end{figure}

In~\cite{Samboy}, we showed that the potential well corresponding to 
the $69s+72s$ asymptote of the $0_g^+$ symmetry curves near Rb $np+np$ 
was composed of five nearby asymptotes. We do not review the detailed 
treatment here, but highlight the significant asymptotes 
in Fig.~\ref{fig:eigvexPLUS}(a) (left panel). 
We note that the composition of this well is due to the strong 
dipole mixing between the five highlighted states, 
which lead to the well having not only $ss'$ character, 
but also $pp'$ character. In the right panel of 
Fig.~\ref{fig:eigvexPLUS}(a), we illustrate the same information
for the well correlated to the $69p_{1/2}+70p_{1/2}$ asymptote
near the $70s+70s$ asymptote of the $0_g^+$ symmetry. Again, we find
that the states contributing the most to the existence of this
well correspond to asymptotes above $69p_{1/2}+70p_{1/2}$.

%%%%%%%%%%%%%%%%%%%%%%%%

%%%%%%%%%%%%%%%%%%%%%%%%%%%%%

In Fig.~\ref{fig:eigvexPLUS}(b), we depict the 
$\ell$-mixing leading to the potential wells in Fig.~\ref{fig:eigvexPLUS}(a):
this is given by the $|c_j(R)|^2$ coefficients. For both wells near $np+np$ 
and $ns+ns$, the molecular basis states $|j\rangle$ mixed by
the dipole and quadrupole interactions correspond to asymptotes that lie
above the asymptote of the wells.
In the case of $70p+70p$, the $69s+72s$ molecular level couples strongly to both the 
$69p+71p$ states (above) and the $68p+72p$ states (below). However, 
the relative energy differences between the asymptotes results in a much stronger interaction
between $69s+72s$ and the $69p+71p$ states than with the $68p+72p$ states.
This is why there is little contribution from the $68p+72p$ states in the formation of the well.
In the case of $ns+ns$, the states directly below the well correlated to the $69p_{1/2}+71p_{1/2}$
asymptote are $68p+71p$ states. In general, the strength of quadrupole coupling between $np$ states
and $n'p$ states is very weak. Combining this with the large spacing between the asymptotic energy
levels results in minimal contributions from the $68p+71p$ states in the formation of the well.

The insignificant contributions of the states below the potential wells in both 
cases indicate little chance of predissociation to these lower asymptotes, and hence the 
macrodimers should be long-lived (limited only by the lifetime of the 
Rydberg atoms). In~\cite{Samboy}, we showed that the nonadiabatic coupling 
between the $69s+72s$ curve and the curves immediately below were very small, 
leading to metastable macrodimers. For the wells near the $ns+ns$ asymptotes,
we reach the same conclusion, {\it i.e.} metastable macrodimers with lifetime
limited by that of the Rydberg atoms.

%%%%%%%%%%%%%%%%%%%%%%%%%%%%%%%%%%%%%%%%%%%%%%%%%%%%%%%%%%%%%%%%%%%%%%%%%%%%

\subsection{Photoassociation}
\label{subs:PA}

\begin{figure}
	\centering
		\includegraphics[width=4.5in]{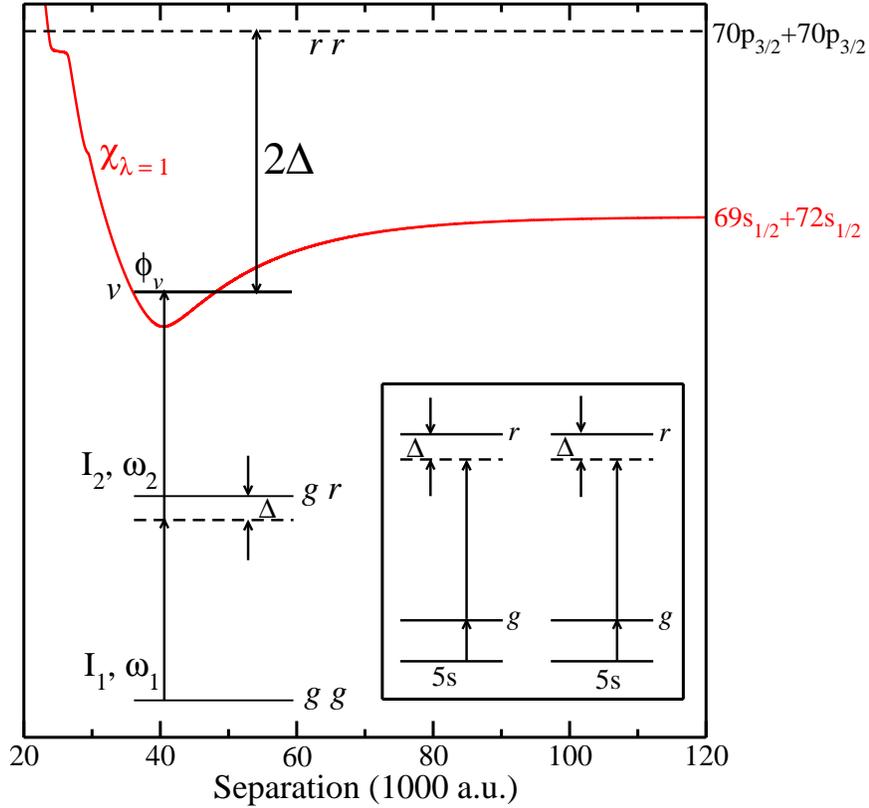}
	\caption{(Color online) Our proposed two-photon photoassociation scheme 
	         for the formation of $70p+70p$ rubidium Rydberg macrodimers. 
	         The ground state atoms are populated to a bound 
	         level inside the well by dual lasers, each of which is red-detuned from the 
	         resonance signal of the $70p_{3/2}+70p_{3/2}$ molecular Rydberg state. 
	         Inset: Each atom is initially excited to an intermediate Rydberg state, which
	         is considered to be the ``ground'' state in our discussion (see text).
	         We note that the single atom detuning levels are not to scale.
           }
	\label{fig:PAscheme}
\end{figure}

Exciting two ground state atoms into a bound level via photoassociation 
(PA) will allow us to probe the different electronic characters mentioned 
in~\ref{subs:Levels}. In the following treatment, we describe a PA scheme 
for the formation of macrodimers bound by the highlighted well of the 
$0_g^+$ symmetry of the $np+np$ asymptote (see Fig.~\ref{fig:eigvexPLUS}(a)
left panel). We assume that the ground state atoms 
are first excited to intermediate Rydberg states 
(treated as the ``ground'' states)
so that the coupling to higher Rydberg states is enhanced. 
%The ``ground'' state atoms
%are selected so that their potential curves
%are asymptotically flat in the $R$ region of the potential 
%well.
Since the bound states of this 
particular well have electronic character that is mostly $\ket{ns;n's}$ and 
$\ket{np;n'p}$, we assumed two possible ``ground" states
making transitions to different $\ell$-character in the well:  $40p_{3/2}+40p_{3/2}$ 
to the $ss'$ components, and $41s+41s$ to the $pp'$ components. 
For simplicity, we choose intermediate states near $n\sim 40$ 
(see inset of figure~\ref{fig:PAscheme}) because the 
electronic potential curves of these states are asymptotically flat in the 
$R$ region of the potential well we wish to populate (see~\cite{Samboy}); 
this greatly simplifies the calculations for the PA rates. 
However, our calculations can of course be modified to fit other experimental parameters 
and conditions.

Figure~\ref{fig:PAscheme} shows a schematic for our proposed two-photon 
formation mechanism
for the case of doubly-excited $np$ atoms.
The macrodimers we predict could be realized and identified spectroscopically
by red-detuning the excitation lasers from the
resonant frequency of the $70p_{3/2}+70p_{3/2}$ molecular Rydberg level.

The PA rate ${\sf K}_v$ for two atoms into a bound level $v$ can be 
calculated~\cite{Robin98} using
\begin{equation}
   {\sf K}_v \propto I_1 I_2\left|\langle \phi_v|\langle \chi_{\lambda}|e^2r_1r_2|
                           \chi_g \rangle |\phi_g\rangle \right|^2 ,
   \label{eq:PARate}
\end{equation}
where $I_1$ and $I_2$ are the intensities of laser 1 and 2, $|\phi_v(R)\rangle$ 
and $|\chi_{\lambda}(R)\rangle$ are the radial and electronic wave functions 
inside the well, respectively, $|\phi_g(R)\rangle$ and $|\chi_g(R) \rangle$ are 
the radial and electronic wave functions of the ground state, respectively, 
and $r_i$ and $e$ are the location and charge of the electron $i$.
%%%%%%%%%%%%%%%%%%%%%%%%%%%%%%%%%%%%%%%%%%%%%%%%%%%%%%%%%%%%%%%
%%%%%%%%%%%%%%%%%%%%%%%%%%%%%%%%%%%%%%%%%%%%%%%%%%%%%%%%%%%%%%%%
Using the expression (\ref{eq:electronic-state}) for $|\chi_{\lambda}(R) \rangle$,
and assuming that $|\chi_g\rangle$ is independent of $R$ (corresponding to 
a flat curve), we can rewrite (\ref{eq:PARate}) as
\begin{equation}
 {\sf K}_v \propto I_1 I_2 \sum_j \left|(d_1d_2)_j\right|^2 \left|\int_0^{\infty}\!\!\!\!
 dR\phi_v^{*}(R)c_j^{*}(R)\psi_g(R)\right|^2 ,
\label{eq:PAInt}
\end{equation} 
where $d_1=\langle n_j\ell_j|er_1|n_g\ell_g\rangle$ and 
$d_2=\langle n'_j\ell'_j|er_2|n'_g\ell'_g\rangle$ are the 
electronic dipole moments between electronic states 
$|a_g; a'_g\rangle$ and $|a_j;a'_j\rangle$ for atom 1 
and atom 2, respectively.

\begin{figure}[h]
	\centering
		\includegraphics[width=6in]{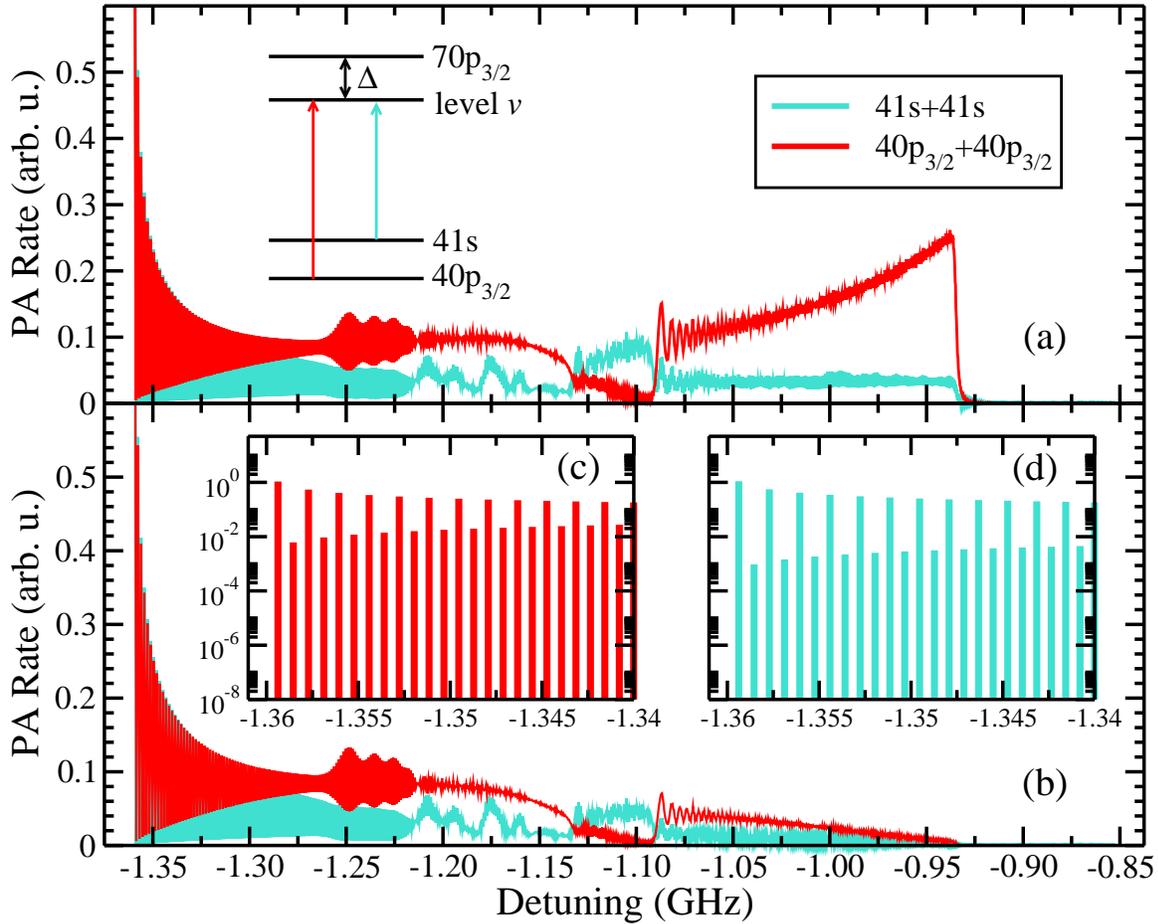}
	\caption{(Color online) PA rate vs. the detuning $\Delta$ from the $70p_{3/2}$ atomic state
	          for (a) a constant radial ground state distribution and (b) a gaussian radial ground
	          state distribution - both on linear scales set to a maximum of 1.
            The $\ket{41s;41s}$ state populates the $pp'$ character in the well (turquoise), 
            and the $\ket{40p_{3/2};40p_{3/2}}$ the $ss'$ (red). 
            Plots (c) and (d) show the rates for the
            deepest levels of the gaussian distribution on a logarithmic scale (see text).
            }
	\label{fig:PABoth}
\end{figure}

In~\cite{Samboy}, we presented PA rate calculations for bound levels in the 
potential well correlated to the $69s+72s$ asymptote
from a flat radial ground state distribution.
The results of the PA rate against the detuning $\Delta$ from the atomic 
70$p_{3/2}$ levels are shown in Fig.~\ref{fig:PABoth}(a).
Since the general expression for the PA rate~\eref{eq:PAInt} is proportional to 
the dipole moments and the laser intensities, our calculated rates are
given in arbitrary units set to a maximum
of one (for the strongest rate starting from $41s+41s$). 
In fact, once a pair of atoms has been
excited to the ``ground" state with fixed laser intensities, the PA rate 
plotted in Fig.~\ref{fig:PABoth} represents the probability of forming a 
macrodimer; the transition from the 5$s$ Rb atoms to the intermediate 
``ground" state atoms can easily be saturated so that the PA process
always starts with a pair of $40p_{3/2}+40p_{3/2}$ or $41s+41s$.
We compare these PA rates to those obtained
if the ground state radial wave function $\phi_g(R)$ is assumed to be
a gaussian centered on $R_e$ with a standard deviation of 14 500 $a_0$ 
(roughly half the FWHM of the potential well) in Fig.~\ref{fig:PABoth}(b).
The choice of a gaussian approximates the wave function obtained by thermally
averaging harmonic oscillator wave functions over a harmonic trapping potential
(\textit{e.g.} in an optical lattice) for both ``ground" state atoms.
In those plots, the PA rates starting from both atoms in $40p_{3/2}$ are 
shown in red and from $41s$ in turquoise, respectively.
The rapid oscillation between large rates for an even bound level 
($v=0,2,4,\dots$) and small rates for odd levels ($v=1,3,5,\dots$) 
gives the apparent envelope of the PA signal. This behavior is due to the oscillatory 
nature of the radial wave functions inside the well $\phi_v(R)$: 
the integral in equation~\eref{eq:PAInt} 
will be near zero for odd wavefunctions. We highlight this for the gaussian
$\psi_g(R)$ distribution of $40p_{3/2}+40p_{3/2}$ in (c) and $41s+41s$ in (d), where we show a 
zoom of the deepest levels $v$ (on a log-scale).

%%%%%%%%%%%%%%%%%%%%%%%%%%%%%%%%%%%%%%%%
%%%%%%%%%%%%%%%%%%%%%%%%%%%%%%%%%%%%%%%%%%%

We see in plots (a) and (b) of Fig.~\ref{fig:PABoth} that the signature of a macrodimer 
would manifest itself by the appearance of a signal starting at $\Delta \sim -0.93$ GHz 
red-detuned from the 70$p_{3/2}$ atomic level, and ending abruptly at $\sim -1.36$ GHz. 
The shape of the signals indicate that the rates can reveal details of the $\ell$-mixing 
inside the potential well; for example, both plots show that the $\mathsf{K}_v$ mimic the 
probabilities $|c_j(R)|^2$ shown in figure~\ref{fig:eigvexPLUS}(b). The progressive 
decrease of the $41s$ signal beginning at $-$1.36 GHz, followed by sharp increases 
between $-$1.22 and $-$1.16 GHz correspond to the slow decreases of the $pp'$ components 
between $R\sim$ 40 000 $-$ 50 000 $a_0$ and their sharp increases around 
$R\sim$ 33 000 $-$ 35 000 $a_0$. For the $40p_{3/2}$ signal, the major feature common 
to both (a) and (b) is the significant drop in ${\sf K}_v$ between $-1.13$ and $-1.09$ GHz, 
which mirrors the decrease in the $ss'$ states between $R\sim$ 52 000 $-$ 55 000 $a_0$ 
in Fig.~\ref{fig:eigvexPLUS}(b).
As noted in~\cite{Samboy}, this range of frequency with a noticeable
drop in the PA rate could serve as a switch to excite or not excite a macrodimer,
depending on the ``ground'' state being used.
We also note that in both cases, the signals 
for the $40p_{3/2}+40p_{3/2}$ ``ground'' state are higher overall across the $R$ regime
of the well, with a few exceptions. This indicates that despite the presence of the 
$pp'$ character, this well is still largely composed of $ss'$ character.

As expected, the gaussian ground state distribution shares many qualitative features 
with the constant ground state distribution. 
%although quantitatively, the gaussian rate 
%signals are much larger. 
We considered both ground state distributions having populations 
in the $R$ range of $\sim$ 30 000 to 70 000 $a_0$. Normalizing both ground state 
distributions over this range yielded slightly larger signal rates in the deepest part of the
well for the gaussian, corresponding to its peak. However, 
%significantly higher probabilities for the gaussian 
%distribution (save for perhaps at the tails of the Gaussian), resulting in much higher rates.
the major difference between Fig.~\ref{fig:PABoth}(a) and (b) is noticeable in both the 
$40p_{3/2}+40p_{3/2}$ and the $41s+41s$ rate signals between $\Delta\sim -1.09$ and $-$0.85 
GHz. Whereas the uniform distribution shows a steady $41s+41s$ signal
and a steady \textit{increase} of the $40p_{3/2}+40p_{3/2}$ signal, the gaussian distribution 
shows both signals rapidly decreasing. As we chose to center the gaussian at the minimum of 
the well, the decrease in both signals obviously corresponds to the decreasing probability 
of the gaussian distribution in this $R$ regime (\textit{i.e.} the tails). 
If one wanted to take advantage of the large, isolated $69s72s$ character at higher $R$, 
this could easily be accomplished by recentering the gaussian appropriately.
%%%%%%%%%%%%%%%%%%%%%%%%%%%%%%%%%%%%%%%%%%%%%

\section{Conclusion}
\label{sec:conc}

We have presented long-range interaction potential curves for the $0_g^+$, $0_u^-$, and $1_u$ 
symmetries of doubly excited $ns$ and $np$ Rydberg atoms and have demonstrated the 
existence of potential wells between these excited atoms. These wells
are very deep and very extended, due to the strong $\ell$-mixing
between the various electronic Rydberg states. These wells are robust against small 
electric fields and support several bound vibrational levels, separated by a few
MHz, which could be detected in spectroscopy experiments. 
The macrodimers corresponding to these 
bound vibrational levels are stable with respect to predissociation and have lifetimes
limited only by the Rydberg atoms themselves. These macrodimers could be realized through
population of the vibrational energy levels by photoassociation, resulting in a detectable
signal that could be used to probe the various $\ell$-character of the potential well.

In conclusion, we note that the detection of such extended dimers could facilitate
studies in a variety of areas. For example, the effect of retardation on the
interaction at very large separation, which becomes important if the photon
time-of-flight between the atoms is comparable to the classical orbital period
of a Rydberg electron around its core~\cite{macro-old}, could potentially be
probed experimentally. Another example relates to chemistry of molecules
with high internal energy; a third atom approaching a macrodimer could quench
its internal state to lower levels, or could potentially react with the molecule 
at very large distances and create a new product such as {\it trilobite} or {\it butterfly} 
Rydberg molecules~\cite{trilobites}. Finally, as mentioned in the introduction,
Rydberg atoms are being investigated intensively for quantum information 
processing, {\it e.g.} using the blockade mechanism~\cite{lukin01}, and the
possibility of frequency ranges where the PA rate is strong or weak 
due to the $\ell$-character mixing could potentially be used as a quantum 
mechanical switch. These few examples illustrate some possible applications
of macrodimers. 

\ack
The work of N.S. was supported by the National Science Foundation, 
and the work of R.C. in part by the Department of Energy, Office of 
Basic Energy Sciences.  

\providecommand{\newblock}{}

\end{document}